\documentstyle[11pt,emp1,epsfig]{article}

\def\be{\begin{equation}}
\def\ee{\end{equation}}
\def\bea{\begin{eqnarray}}
\def\eea{\end{eqnarray}}

\begin{document}
\renewcommand{\thefootnote}{\fnsymbol{footnote}}
\vspace*{4cm}
\title{ELASTIC MESON PRODUCTION -- FACTORISATION AND GAUGE INVARIANCE
\footnote{presented by A. Hebecker at the Rencontres de Moriond, 
"QCD and High Energy Hadronic Interactions", Les Arcs, March 1998
}}

\author{A. HEBECKER and P. V. LANDSHOFF}

\address{DAMTP, Cambridge University, Cambridge CB3 9EW, England}

\maketitle\abstracts{
The factorisation of the hard amplitude for exclusive meson production in
deep inelastic scattering is considered in the framework of a simple model.
It is demonstrated explicitly how gauge invariance ensures the cancellation
of non-factorising contributions.}

\section{Introduction}
Recent HERA measurements \cite{exp} of exclusive vector meson 
electroproduction at high photon energy and virtuality open new 
possibilities for the investigation of the interplay of hard and soft 
physics. The original crude calculations based on a simple model of two 
gluon exchange \cite{ad} have since been refined by several authors 
\cite{pc}. Recently, the mechanism of factorisation of the calculable hard 
amplitude and the nonperturbative component, given by the non-diagonal 
parton densities and the meson wave function, has been discussed in a rather 
general framework \cite{cfs}. 

It is the purpose of this talk to demonstrate explicitly, in the framework 
of a simple model, how gauge invariance ensures the cancellations required 
for the factorisation of the hard part of the amplitude \cite{hl}.

In the simplest diagrammatic description, the incoming photon splits into a 
quark antiquark pair which then interacts with the target via two gluon 
exchange and finally forms the outgoing vector meson. We focus on the 
problems associated with the vertex that couples the outgoing quarks to the 
vector meson. In the case of light flavours, this vertex is nonperturbative 
and so our knowledge of it is far from complete, though its analytic 
structure is known \cite{elop}. If we neglect spin, it is a function 
$V(u,v)$ of the squared 4-momenta on the two quark legs, $u$ and $v$. 
Branch points in each of its variables, $u$ and $v$, are associated 
with ``normal thresholds'' and possibly also ``anomalous thresholds''. We 
show that even in leading power this results in a breakdown of the desired 
diagrammatic factorisation.

However, this does not necessarily imply that the factorisation theorem is 
invalid. For a generic nonperturbative vertex function $V$, the naive two 
gluon exchange calculation is not QCD gauge invariant. To achieve gauge 
invariance, graphs in which either or both of the exchanged gluons couples 
directly into the nonperturbative meson vertex function have to be included.
The question we discuss is whether these additional diagrams restore the 
factorisation theorem.

We are not able to give a definitive answer to this question, because
of the need to introduce two further nonperturbative vertex functions,
and we know as little about these as we do about $V$. However, we report
a calculation based on an explicit simple model in which we find that
the factorisation theorem is indeed restored. This encourages the belief
that its validity may be general.

Our model is a simple one in which all three vertex functions have the 
analytic structure that is expected on general grounds \cite{elop} and in 
which they are related in such a way that the complete leading-power 
amplitude for the exclusive meson production is gauge invariant. We then 
find that the amplitude contains leading contributions from diagrams that 
cannot be separated into a hard production process and soft wave function 
corrections. However, gauge invariance leads to a cancellation 
of the unwanted contributions and leaves the factorisation theorem intact.
The final formula can be obtained from the leading order diagrams by a 
redefinition of the rules for calculating them: basically, this amounts to 
ignoring the branch points in the vertex function.

In Sec.~2, we recall some aspects of the calculation in the case where the 
structure of the nonperturbative vertex function $V$ may be neglected. In 
Sec.~3 we explain the complications that arise when its structure is taken 
into account, introduce a simple model for this structure and analyse its
consequences.

\begin{figure}[ht]
\begin{center}
\vspace*{-.5cm}
\parbox[b]{13.3cm}{\epsfxsize=13.3cm\epsfbox{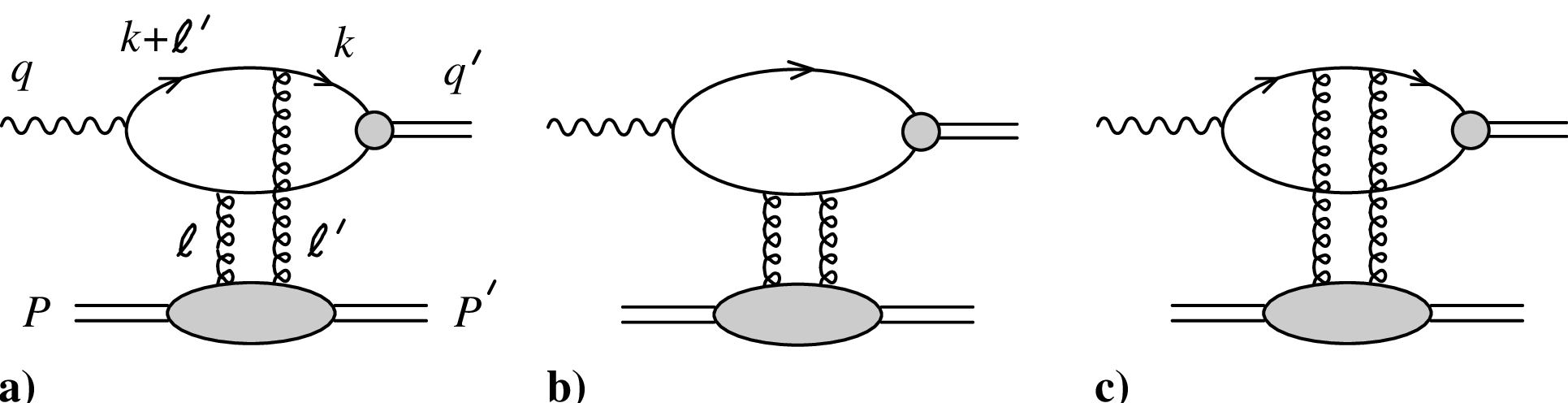}}\\
\end{center}
\refstepcounter{figure}
\label{lo}
{\bf Fig.\ref{lo}} 
The leading amplitude for a structureless meson vertex.
\end{figure}

\vspace*{-.4cm}
\subsubsection*{2. Structureless vertex function}
In this section, we focus on the case of a structureless vertex function 
corresponding to a point-like coupling of the quarks to the outgoing meson
(see Fig.~\ref{lo}). For simplicity, we take the meson to be spinless, and 
also pretend that the photon and the quarks have no spin. The coupling of 
the gluons to the scalar quarks is given by $-ig\,(r_\mu+r_\mu')$, where 
$r$ and $r'$ are the momenta of the directed quark lines, and the coupling 
of the photon is $ie$, where $e$ has the dimensions of mass. We use 
light-cone co-ordinates and work in a frame where $q,q'$ and $P,P'$ have 
zero transverse components and large `$+$' and `$-$' components 
respectively. We  concentrate on the forward production, so that the 
transverse component of the momentum transfer $\Delta=q'-q=\ell-\ell'=P-P'$ 
vanishes, $\Delta_\perp =0$. 

We study the limit $W^2\gg Q^2\gg m_V^2$, where $W^2$ is the $\gamma^*p$ 
energy. The diagrams of Fig.~\ref{lo} are the only ones that contribute in 
leading power when $V$ is constant. The lower bubble in these diagrams in 
principle has a 
complicated structure. However, we shall
assume that the main contribution comes from values of its subenergy 
$\sigma$ that are not too large,
$\sigma =(P-\ell)^2\ll W^2$. Then $\ell _+\ll q_+$ and the dependence of 
the lower bubble on $\sigma$ may be approximated by $\delta (\sigma )$. 
Further, when the upper parts of the diagrams are added together the
main contribution arises from small values of
$\ell_-,\quad\!\!\ell _-\ll P_-$, so that also $\ell '_-\ll P_-$ 
and $\ell ^2\sim\ell^{\prime 2}\sim-\ell_\perp^2$. This may be seen most 
simply by calculating the imaginary part of the amplitude, where the left-%
most of the two quarks to which $\ell'$ is attached is on shell.
Hence the important part of the lower bubble effectively has the structure
\be
F^{\mu\nu}(\ell,\ell',P)\approx\delta(P_-\ell_+)\,F(\ell_\perp^2)\,P^\mu 
P^\nu\,,\label{fd}
\ee
which is defined to include both gluon propagators and all colour factors. 
A similar expression was found by Cheng and Wu \cite{cw} in
a tree model for the lower
bubble, though we do not need to restrict ourselves to such a simple model.
We assume that $F$ restricts the gluon momentum to be soft, 
$\ell_\perp^2\ll Q^2$. In the high energy limit it suffices to calculate 
\be
M=\int\frac{d^4\ell}{(2\pi)^4}T^{\mu\nu}F_{\mu\nu}\approx\int
\frac{d^4\ell}{4(2\pi)^4}T_{++}F_{--}\,,\label{it}
\ee
where 
\be
T^{\mu\nu}=T^{\mu\nu}(\ell,\ell',q)=T^{\mu\nu}_a+T^{\mu\nu}_b+T^{\mu\nu}_c
\ee
is the sum of the upper parts of the diagrams in Fig.~\ref{lo}. 

\def\half{{\scriptsize\frac{1}{2}}}
The lower amplitude $F_{\mu\nu}$ in the diagrams of Fig.~\ref{lo} is
symmetric with respect to the two gluon lines. This symmetry of the lower 
amplitude allows us to replace the properly-symmetrised upper amplitude 
$T_{\mbox{\scriptsize sym}}^{\mu\nu}$ with the unsymmetrised amplitude
$T^{\mu\nu}$ corresponding to the sum of the diagrams in Fig.~\ref{lo}. The
symmetrised amplitude is
\be
T_{\mbox{\scriptsize sym}}^{\mu\nu}(\ell,\ell',q)=\half [
T^{\mu\nu}(\ell,\ell',q)+ T^{\nu\mu}(-\ell',-\ell,q)]\,.
\ee
The two exchanged gluons together must form a colour singlet and so, at 
least for our calculation that takes account only of the lowest order in 
$\alpha_S(Q^2)$, the symmetrised
amplitude $T_{\mbox{\scriptsize sym}}^{\mu\nu}$ 
satisfies the same Ward identity as for two photons:
\be
T_{\mbox{\scriptsize sym}}^{\mu\nu}(\ell,\ell',q)\ell_\mu \ell_\nu'=0\,.
\label{w1}
\ee
Writing this equation in light-cone components and setting $\ell_\perp=
\ell_\perp'$, we see that for the small values of $\ell _-$, $\ell '_-$,
$\ell_+$ and $\ell_+'$ that we need,
\be
T_{\mbox{\scriptsize sym,}++} \sim\ell_\perp^2\,
\ee
for $\ell_\perp^2\to 0$. Here we have used the fact that the tensor 
$T_{\mbox{\scriptsize sym}}^{\mu\nu}$, which is built from $\ell'$, $\ell$ 
and $q$, has no large `$-$' components. The $\ell_-$ integration makes this 
equation hold also for the original, unsymmetrised amplitude: 
\be
\int d\ell_-T_{++} \sim\ell_\perp^2\,.\label{w2}
\ee
This is the crucial feature of the two-gluon amplitude that will simplify 
the calculation and lead to the factorising result of the next section. 

It is convenient to begin with the contribution from diagram a) of 
Fig.~\ref{lo} to the $\ell_-$ integral of $T_{++}$, which is required in 
(\ref{it}):
\be
\int d\ell_-T_{a,++}=-4eg^2q_+\int\frac{d^4k}{(2\pi)^3}\,\frac{z(1-z)}
{N^2+(k_\perp+\ell_\perp)^2}\,\frac{V(k^2,(q'-k)^2)}{k^2(q'-k)^2}\,.
\label{ta}
\ee
Here $N^2=z(1-z)Q^2$, $z=k_+/q_+$ and the condition $\ell_+=0$, enforced 
by the $\delta$-function in Eq.~(\ref{fd}), has been anticipated. 

Now $\int d\ell_-T_{b,++}$ and $\int d\ell_-T_{c,++}$ each carry no 
$\ell_\perp$ dependence. So to ensure the validity of Eq.~(\ref{w2}) the sum 
of the three diagrams must be 
\be
\int d\ell_-T_{++}=4eg^2q_+\int\frac{d^4k}{(2\pi)^3}z(1-z){\cal N}
\frac{V(k^2,(q'-k)^2)}{k^2(q'-k)^2}\,,
\label{intit}
\ee
where
\be
{\cal N}=\Big [\frac{1}{N^2+k_\perp^2}-\frac{1}
{N^2+(k_\perp+\ell_\perp)^2}\Big ]\sim \frac{\ell_\perp^2}{N^4}\,.
\label{intitt}
\ee
We have used the softness of the wave function, which results in the 
dominant contribution to the integral arising from values of 
$k_\perp^2\ll Q^2$, and the rotational symmetry, which makes 
$k_\perp\cdot\ell_\perp$ integrate to 0. Note the $1/Q^4$ behaviour obtained 
after a cancellation of $1/Q^2$ contributions from the individual diagrams. 
This cancellation, which is closely related to the well-known effect of 
colour transparency \cite{ct}, has been discussed in \cite{ad} in the 
framework of vector meson electroproduction. 

Introduce the light-cone wave function of the meson
\be
\psi(z,k_\perp^2)=-\frac{iq'_+}{2}\int dk_-dk_+\,
\frac{V(k^2,(q'-k)^2)}{(2\pi)^4
k^2(q'-k)^2}\,\delta (k_+-zq'_+).\label{wv}
\ee
The final result following from Eqs.~(\ref{it}) and (\ref{intit}) is a 
convolution of the production amplitude of two on-shell quarks and the 
light-cone wave function:
\be
M=ieg^2W^2\left(\int\frac{d^2\ell_\perp}{2(2\pi)^3}\ell_\perp^2
F(\ell_\perp^2)\right)\int dz\int d^2k_\perp\frac{z(1-z)}{N^4}
\psi(z,k_\perp^2)\,.\label{lot}
\ee
This corresponds to the $O(\ell_\perp^2)$ term in the Taylor expansion
of the contribution Eq.~(\ref{ta}) from Fig.~\ref{lo}a).

\subsubsection*{3. Factorisation for a simple model wave function}
In the previous section we have supposed that the vertex function $V$ 
has no structure. We used the gauge invariance of the amplitude to 
argue that there is a cancellation among the three diagrams of 
Fig.~\ref{lo}a). Of course, we may instead obtain the same result by 
explicit calculation of each of the three diagrams. This may be done most 
simply by calculating their imaginary parts and using the known fact that 
the complete amplitude must be pure-imaginary when its energy dependence is 
$(W^2)^{1.0}$. Alternatively, the contributions to the amplitude itself may 
be calculated by doing the $k_-$ integration, completing the integration 
contour with an infinite semicircle and taking appropriate pole residues. 

Consider Fig.~\ref{lo}b) for example. In either method of calculation,
if $V$ has no structure the upper quark line gets put on shell.
However, if we now take account of the known structure of $V$, there is
an additional contribution to the imaginary part corresponding to cutting
the graph through the vertex function --- the $k_-$ integration would
need to take account also of branch points of $V$, not just the poles
of the propagators. The gauge-invariance argument breaks down because
the set of diagrams by itself is no longer gauge invariant: one must
add to it diagrams where either or both of the gluons couples directly
into the vertex function.

\begin{figure}[ht]
\begin{center}
\vspace*{-.2cm}
\parbox[b]{11.1cm}{\epsfxsize=11.1cm\epsfbox{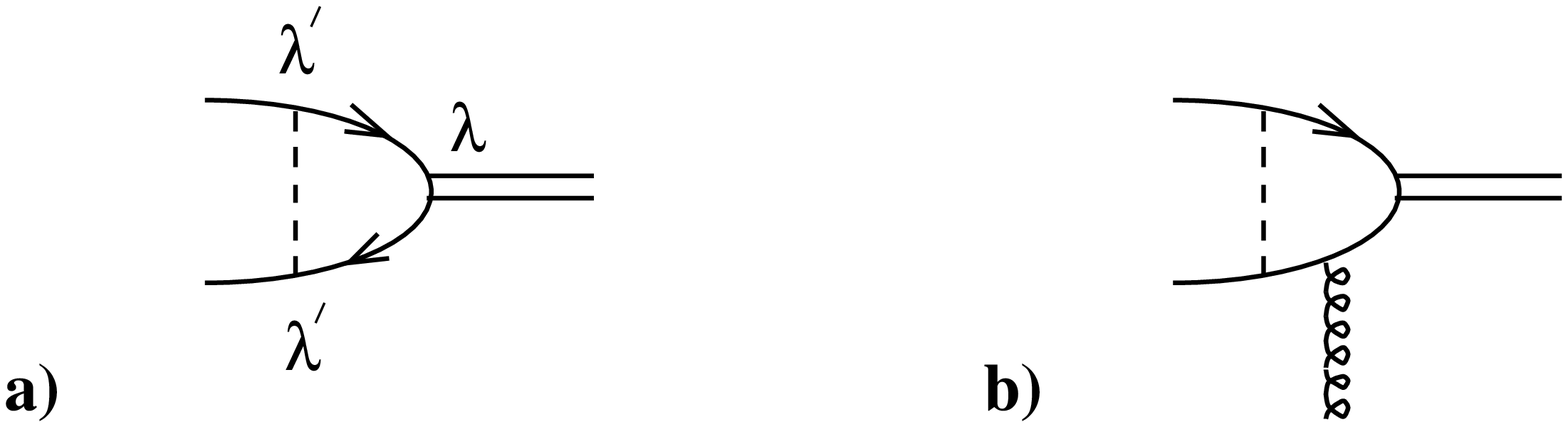}}\\
\end{center}
\refstepcounter{figure}
\label{ver}
{\bf Fig.\ref{ver}a)} Model for the nonperturbative vertex function $V$.
{\bf b)} Diagram with a gluon coupling into the vertex.
\end{figure}

In order to study this, we use the simplest model for the quark-quark-meson 
vertex function $V(u,v)$ that incorporates at least part of its known 
branch-point structure. It is the purely nonperturbative vertex function 
that we are modelling: it goes to zero suitably rapidly when either of the 
squared 4-momenta $u$ or $v$ of the quarks becomes large. We do not consider 
its perturbative tail, which would be obtained by exchanging a perturbative 
gluon between the quarks. It is a familiar notion \cite{elop} that the 
correct {\it analytic} properties of nonperturbative amplitudes are those 
corresponding to Feynman graphs, even though the {\it numerical} values of 
such graphs have no physical significance. In order to model the vertex 
function $V$, therefore, we use the simple Feynman graph of 
Fig.~\ref{ver}a), where the line that joins the quarks is a scalar (like 
the quarks themselves in our simple calculations) which couples to them with 
strength $\lambda'$ and where the right-hand internal vertex is taken to be 
a constant $\lambda$. This model has branch points in each of the variables 
$u$ and $v$, and it has the appropriate softness. 

\begin{figure}[ht]
\begin{center}
\vspace*{-.5cm}
\parbox[b]{4.8cm}{\epsfxsize=4.8cm\epsfbox{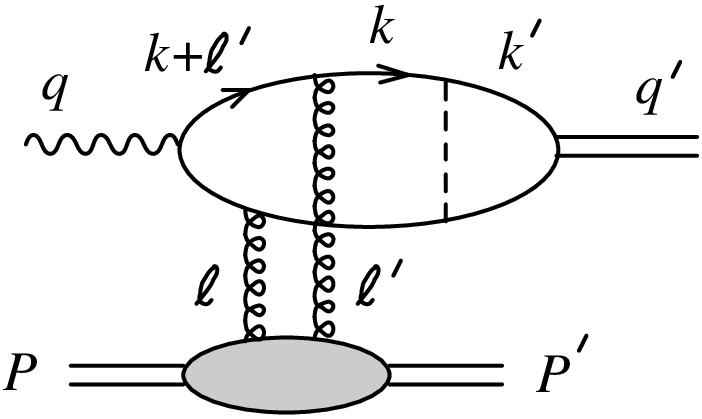}}\\
\end{center}
\refstepcounter{figure}
\label{nlo}
{\bf Fig.\ref{nlo}} 
Diagram for meson production with the vertex modelled by scalar 
particle exchange.
\end{figure}

\begin{figure}[ht]
\begin{center}
\vspace*{-.5cm}
\parbox[b]{13.3cm}{\epsfxsize=13.3cm\epsfbox{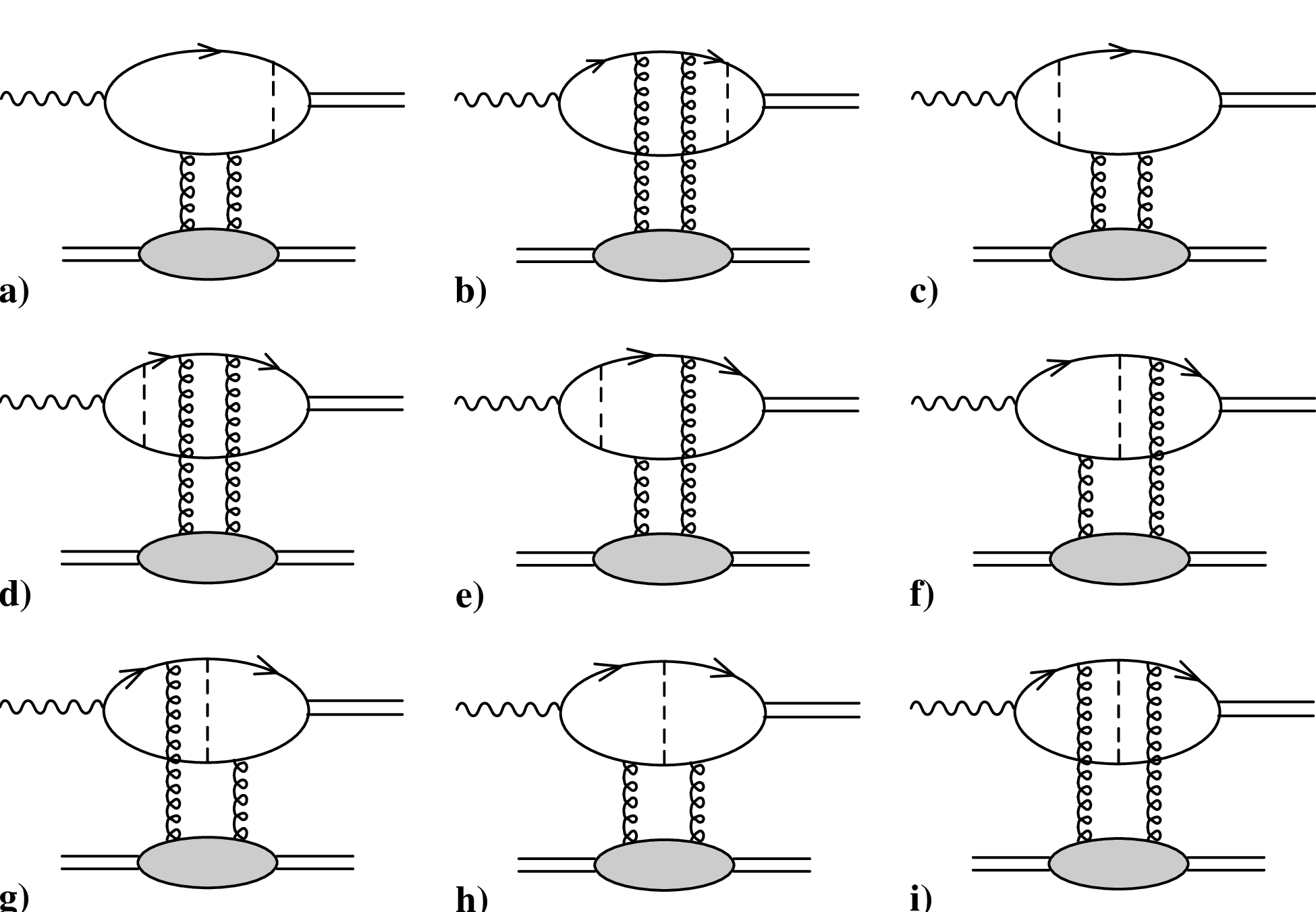}}\\
\end{center}
\refstepcounter{figure}
\label{rest}
{\bf Fig.\ref{rest}} 
The remaining diagrams contributing to meson production within the above 
simple model for the meson wave function.
\end{figure}

When the vertex function of Fig.~\ref{ver}a) is used in Fig.~\ref{lo}a), 
we obtain Fig.~\ref{nlo}. The expression for the upper part of the diagram 
is Eq.~(\ref{ta}) with 
\be
V(k^2,(q'-k)^2)=\int\frac{d^4k'}{(2\pi)^4}\,\frac{i\lambda\lambda'^2}{k'^2
(q'-k')^2(k-k')^2}\,.\label{tf2v}
\ee
The diagram of Fig.~\ref{nlo} by itself gives no consistent description of 
meson production since it lacks gauge invariance. This problem is not cured 
by just adding the two diagrams \ref{lo}b) and c) with the blob replaced by 
the vertex of Fig.~\ref{ver}a). It is necessary also to include 
diagrams where a gluon is coupled into the vertex (see, e.g., 
Fig.~\ref{ver}b)). Furthermore, diagrams where both gluons couple into 
the vertex have to be included. The complete set of these additional 
diagrams is shown in Fig.~\ref{rest}.

The same gauge invariance arguments that lead to Eq.~(\ref{w2}) apply to 
the sum of all the diagrams in Figs.~\ref{nlo} and \ref{rest}. Therefore, 
the complete result for $T_{++}$, which is now defined by the sum of the 
upper parts of all these diagrams, can be obtained by extracting the 
$\ell_\perp^2$ term at leading order in $W^2$ and $Q^2$. Such a term, with 
a power behaviour $\sim \ell_\perp^2W^2/Q^4$, is obtained from the diagram 
in Fig.~\ref{nlo} (see Eqs.~(\ref{ta}) and (\ref{tf2v})) by expanding around 
$\ell_\perp=0$. It can be demonstrated that none of the other diagrams 
gives rise to such a leading-order $\ell_\perp^2$ contribution (see 
\cite{hl} for more details). 

The complete answer is given by the $\ell_\perp^2$ term from the Taylor 
expansion of Eq.~(\ref{ta}). The amplitude $M$ is precisely the one of 
Eqs.~(\ref{lot}) and (\ref{wv}), with $V(k^2,(q'-k)^2)$ given by 
Eq.~(\ref{tf2v}). We have also checked the correctness of this simple 
factorising result by explicitly calculating all diagrams of Fig.~\ref{rest}.

\subsubsection*{5. Conclusions}
The mechanism of factorisation in exclusive meson production has been 
analysed in the framework of a simple scalar model. In this model the meson 
is formed by two scalar quarks interacting via the exchange of a scalar 
boson. From a calculation of
all  contributing diagrams within the restriction of two-%
gluon exchange the following picture emerges.

The complete result contains leading contributions from diagrams that 
cannot be factorised into quark-pair production and meson formation. 
Nevertheless, in Feynman gauge the answer to the calculation
can be anticipated by looking only at one 
particular factorising diagram. The reason for this simplification is gauge 
invariance. In the dominant region where the transverse momentum $\ell_\perp$ 
of the two $t$-channel gluons is small,  gauge invariance requires the 
complete quark part of the amplitude to be proportional to $\ell_\perp^2$. 
The leading $\ell_\perp^2$ dependence comes exclusively from one 
diagram. Thus, the complete answer can be obtained from this particular 
diagram, which has the property to factorise explicitly if the two quark 
lines are cut. The resulting amplitude can be written in a factorised form. 

It was our intention to demonstrate the above mechanism of gauge invariance 
induced cancellations in as simple and as explicit a way as possible.
Our analysis supplements the
otherwise much more general and complete discussion of \cite{cfs} 
by making it more explicit and by handling the known structure of
the nonperturbative meson vertex function.

\end{document}